\documentclass{aa}

\usepackage{graphicx}
\usepackage{subfigure}
\usepackage{txfonts}

\newcommand{\sax}{{\it BeppoSAX}}
\newcommand{\swf}{{\it Swift}}

\newcommand{\rxte}{{\it RXTE}}
\newcommand{\xmm}{{\it XMM-Newton}}

\newcommand{\asca}{{\it ASCA}}
\newcommand{\euve}{{\it EUVE}}
\newcommand{\mrk}{Mrk~421}

\newcommand{\X}{{\it XMM~}}
\newcommand{\FULL}{{\it FULL~}}

\def\nh{$N_{\rm{H}}$~}

\begin{document}

\title{Signatures of synchrotron emission and of electron acceleration in the X-ray spectra
of \mrk}

\author{
      A.~Tramacere\inst{1}
     \and F.~Massaro\inst{2}
     \and A.~Cavaliere\inst{2}
}

\institute{
Dipartimento di Fisica, Universit\`a di Roma La Sapienza,
Piazzale A. Moro 2, I-00185 Roma, Italy 
\and Dipartimento di Fisica, Universit\`a di Roma Tor Vergata,
Via della Ricerca scientifica 1, I-00133 Roma, Italy 
}

\offprints{andrea.tramacere@roma1.infn.it}
\date{Received ....; accepted ....}

\markboth{A. Tramacere et al.:}
{A. Tramacere et al.:}

\abstract{BL Lac objects undergo strong flux variations involving considerable changes in
their
spectral shapes.
We specifically investigate the X-ray spectral evolution of \mrk~ over a time span of
about nine years.
}
{ 
We aim at statistically describing and physically understanding the large spectral 
changes in X rays observed in \mrk~ over this time span.
}
{
We perform a homogeneous spectral analysis of a wide data set including
archived observations with \asca, \sax, \rxte, as well as published and unpublished
\xmm~ data. The presence of uncertainties is taken into account in our correlation
analysis. The significance of the correlations found  and possible spurious effects are
studied with Monte Carlo simulations. 
}
{
We find that the \mrk~  spectral energy distribution (SED) has a lower peak 
at energies that vary in the range, 0.1-10 keV while its X-ray spectrum is definitely curved.
Parameterizing the X-ray spectra with a log-parabolic model, we find a positive
correlation between the position and the height of the SED peak. 
In addition, we find a negative trend of the spectral curvature parameter vs. the SED
peak energy.
}
{
We show that these relations between the spectral parameters are consistent with
statistical or stochastic acceleration of the emitting particles, and provide insight into
the physical processes occurring in BL Lac nuclei.
}

\authorrunning{A. Tramacere et al.}
\titlerunning{Signatures of synchrotron emission and of electron acceleration in the X-ray
  spectra of \mrk}

\maketitle


\section{Introduction}
In the  popular unification scenario of the Active Galactic  Nuclei (AGNs) BL 
Lac objects are understood as
sources in which a relativistic jet is produced by the central engine and points close to
our line  of sight, see Antonucci (1993) and Urry \& Padovani  (1995).
These objects are marked by featureless spectra and  a high flux from the radio to the 
$\gamma$-ray band, endowed with strong variability and with high polarisation observed in
the optical band.  
In fact, the emission from BL Lacs constitutes one of the best examples of direct
non-thermal radiation from particles not in thermal equilibrium with photons nor among
themselves.
So these sources are keenly interesting to investigate particles acceleration mechanisms
in AGNs.

On a close look, their Spectral  Energy Distribution  (SED) appears as double-peaked. The 
lower
energy component is widely interpreted as synchrotron emission from highly relativistic 
electrons with Lorentz factors $\gamma$ in excess of $10^2$; often it peaks at frequencies
in the IR to the X-ray band. The actual position of this peak 
has been suggested by Padovani \& Giommi (1995)  as a marker for a  
classification;   they define LBL (Low energy peaked BL Lac) objects with the first bump
in the IR--to--optical band, as opposed to the HBL (High energy peaked BL Lac) that peak
in the UV--X-ray band. 
According to the Synchrotron Self Compton (SSC) emission mechanism,   
the high frequency  bump is to be attributed to inverse
Compton scattering of synchrotron photons by the same population of
relativistic electrons that produce the  synchrotron emission (Jones, O'Dell \&
Stein   1974, Ghisellini \& Maraschi 1989). Both emissions are produced in a 
relativistic outflow
with a bulk Lorentz factor $\Gamma \approx 10$;  when observed at an angle $\theta$ they
are subjected to the effects of a beaming factor $\delta = 1/(\Gamma  (1 - \beta \cos
\theta ))$.

With its redshift $z$ = 0.031, \mrk~  is among the closest and best studied HBL.
In fact, it is one of brightest BL Lac objects in the UV and in the X-ray bands,
observed in $\gamma$ rays by EGRET (Lin et al. 1992); it was also the first extragalactic
source detected at TeV energies in the range 0.5-1.5 TeV by the Whipple telescopes (Punch
et al. 1992, Petry et al. 1996).

The source is classified as HBL because its synchrotron emission peak ranges  from a
fraction of a keV to several keVs. In fact, its flux changes go along with strong spectral
variations (Fossati  et al.  2000a).
The spectral shape generally  exhibits  a  marked  curvature, well  described  by  a
log-parabolic model (Landau et al. 1986).

We present here a new study of the time-variable properties of its 
SED based on the full collection of \xmm~ data taken with EPIC CCD
cameras, sensitive in the energy range 0.3--10 keV;  we also use 
previous X-ray observations with \asca, \sax, \rxte, \euve~ to cover an overall time span
of nine years. In particular, \xmm~ observed \mrk~ many times, and a large amount of data
has been collected. Some of these observations have been recently analysed by other
authors, but a
full report of all observations between May 2000 and November 2005 had not been published
yet.
We perform a new analysis of all \xmm~ observations of \mrk,  particularly  relevant as
its  synchrotron emission peaks in the \xmm~ energy range. 

Specifically, extending the previous work of Tanihata  et al. (2004),  we study
correlations  between the position and the height of SED peak, and interpret them in terms
of signatures of the synchrotron emission. We also investigate a new correlation of the
SED peak with the spectral curvature, and discuss its implications as to the electron
acceleration mechanisms.

This paper is organized as follows.
Sect. 2 describes the data set and our procedure to reduce the \xmm~ data.
Sect. 3 reports our data analysis, including temporal and spectral investigations.
The statistical analysis and the comparison with numerical simulations and theoretical 
interpretations are discussed in Sect. 4 and Sect. 5. 
Our conclusions are given and discussed in Sect. 6.

\section{Data sets and Reductions}
\subsection{\xmm~ observations}
\begin{table}
\caption{\xmm~ Log of observations of Mrk 421.}
\begin{flushleft}
\begin{tabular}{ccccccccccccc}
\hline
\multicolumn{1}{c}{Date} & \multicolumn{4}{c}{EPIC-MOS1} & \multicolumn{4}{c}{EPIC-MOS2}\\
\multicolumn{1}{c}{dd/mm/yy} & \multicolumn{1}{c}{Frame} & \multicolumn{1}{c}{Filter} & \multicolumn{1}{c}{Exp} & \multicolumn{1}{c}{Frame} & \multicolumn{1}{c}{Filter} & \multicolumn{1}{c}{Exp}\\
\hline  
\noalign{\smallskip}
25/05/2000 & FU & Md & 18000 & PW & Md & 24001 \\
01/05/2000 & PW & Md & 37251 & PW & Md & 37250 \\      
13/05/2000 & PW & Md & 46950 & PW & Md & 46951 \\         
14/05/2000 & PW & Md & 41948 & PW & Md & 41948 \\ 
08/05/2001 & PW & Tn & 38408 & PW & Tn & 38409 \\
04/05/2002 & FU & Tn & 38898 & FW & Tn & 39163 \\ 
05/05/2002 & FW & Tk & 19422 & FW & Tn & 19419 \\
04/05/2002 & FU & Tn & 23407 & FU & Tn & 23409 \\
04/05/2002 & FU & Tn & 23416 & FU & Tn & 23410 \\
14/05/2002 & FW & Tn & 71200 & FW & Tn & 71200 \\
14/05/2002 & FW & Tn & 11172 & FW & Tn & 11174 \\ 
15/05/2002 & FW & Tn & 11175 & FW & Tn & 11174 \\ 
01/05/2002 & FW & Tn & 70873 & FW & Tn & 70873 \\
02/05/2002 & FW & Tn & 11173 & FW & Tn & 11172 \\
02/05/2002 & FW & Tn & 11171 & FW & Tn & 11175 \\
01/05/2003 & FU & Md & 40542 & PW & Md & 40923 \\
02/05/2003 & FU & Tk & 8912  & PW & Tk & 8911  \\
07/05/2003 & FU & Tk & 8000  & PW & Tk & 8000  \\ 
14/05/2003 & PW & Tn & 48668 & FU & Tn & 48413 \\
10/05/2003 & PW & Md & 25848 & PW & Tk & 25851 \\
06/05/2004 & PW & Md & 61532 & PW & Md & 61534 \\ 
07/05/2005 & RF & Tn & 9765  & RF & Tn & 9771  \\
07/05/2005 & RF & Md & 9465  & RF & Md & 9470  \\   
09/05/2005 & PW & Tk & 59809 & PW & Tk & 59802 \\ 
\noalign{\smallskip}
\hline
\end{tabular}
\end{flushleft}
(a) FRAME: PW=partial window, FU=fast uncompressed, FW=full window, RF=refresh frame.

(b) FILTER: Tn=thin, Md=Medium, Tk=thick.
\end{table}

\mrk~ was observed by \xmm~ between 25/05/00 and 9/11/05, on more than twenty
occasions, by means of  all EPIC CCD cameras: the EPIC-PN (Struder et al. 2001),
and EPIC-MOS  (Turner et al. 2001), operating in different  modes and
with different filters as described in Table 1.
These data were reduced as follows. 
Extractions of all light curves, source  and background spectra  were done
using the \xmm~ Science  Analysis System (SAS) v6.5.0.  The Calibration
Index File  (CIF) and  the summary file  of the Observation  Data File
(ODF) were  generated using Updated Calibration File (CCF) following the ``User's Guide 
to the \xmm~ Science Analysis Syste''  (Issue 3.1) (Loiseau et  al. 2004) and ``The \xmm~
ABC Guide''  (vers. 2.01)  (Snowden et al.  2004). 
Event files were producted by \xmm~ EMCHAIN pipeline.
The  standard reduction  of the  events list  for MOS  data, was performed
involving subtraction of hot and  dead pixels, removal of events due to
the  electronic noise  and  correction of  event  energies for  charge
transfer losses.

To provide the most conservative screening criteria MOS data files were also filtered to 
include all single to quadruple events (PATTERN $\leq  $ 12) with pulse high rate  in the
range of  500 to 12,000 eV and  with expression FLAG $=$ 0.  Lightcurves  for  every
dataset  were extracted  and  all high-background  time intervals, 
were filtered out by excluding  time  interval contaminated  by  solar  flare signal.
To perform this selection, the count rate in the 10 -- 15 keV band for the entire MOS 
detectors were determined. 
We first discarded only time intervals with a count rate that exceeds 0.35 counts per 
second as indicated in the ``User's Guide  to the \xmm~ Science Analysis System'' (Issue
3.1) (Loiseau et  al. 2004).
However applying this criterion, we noticed that a low background state,
placed between two high neighboring peaks due to a solar flare, 
can even include a residual contamination to the source signal which modifies the spectral
distribution.
We adopted a more conservative selection  for good
time intervals, because  we preferred to have a high quality signal,
excluding  time  ranges  that appeared  contaminated  by  solar flares. 
Then we selected good time intervals by direct inspection far from solar flare peaks and 
without count rate variations on time scales shorter than 500 seconds.
The  TABGTIGEN task of {\it  XMM-Newton} Science  Analysis System (SAS) was used to build
good time intervals.

Photons were extracted from an annular region using
different apertures to minimize  pile-up, which affects MOS data.  The
mean value of  external radius for the annular  region is $40''$.
To filter out pixels affected by significant pile-up, the internal
region was selected by using EPATPLOT task in {\it XMM-Newton} Science
Analysis System (SAS) for each observation.

In FULL WINDOW images, the background spectrum was extracted from a
circular region of the size comparable to the source region, in a
place where visible sources were not present (typically off axis). 
For other observations, in PARTIAL WINDOW images, no regions sufficiently far from the
source for the background extraction were found.
In these cases we used background from blank-field event files (www.sr.bham.ac.uk).
Anyway we estimated that the average X-rays background flux was
always at $\sim$ 1\% level of source flux, resulting in a negligible
contamination in the spectral parameter determination.
The Photon Redistribution Matrix and the Ancillary Region
File were created for each observation, by  using RMFGEN  and ARFGEN tasks of SAS.
A more restricted energy range (0.5--10 keV) was used to account
for possible residual calibration uncertainties.   
To insure the validity  of  Gaussian  statistics,  data were  grouped  by  combining
instrumental channels so that each new bin comprises  40 counts or more.

\subsection{Other X-ray observations}
To study the spectral behaviour of \mrk~ over a wide time interval we also used other
X-ray  observations performed with \asca, \rxte, \euve~ and \sax, with references given
below.
All these observations were analysed in terms of the same spectral model as the \xmm~ ones
homogeneous (described in \S 3), so we have
obtained a  sample of spectral parameters that covers about 9 years of \mrk~ observations
in X rays.

\mrk~ was pointed continuously by  \asca~ for 7 days between 23 April of 1998 and 30 April
 1998, with all its detectors on: two SISs (Stage Imaging Spectrometers) and two GISs (Gas
Imaging Spectrometers).
To extend the \asca~ data base to higher energy, these observations were coordinated with 
\rxte, specifically   with the PCA (Proportional Counter Array; 2--60 keV) and the HEXTE
(High Energy X-ray Timing Experiment; 15--200 keV) instruments on board  \rxte. 
We refer to Tanihata et al. (2001) for the \asca~ data reduction and spectral analysis,
and  to Tanihata et al. (2004) for the the  combination of \rxte~ with \asca. 

\sax~ observed \mrk~ on 7 occasions during 1997, in which all instruments on board were 
operating: LECS (Low Energy Concentrator Spectrometer; 0.1--10 keV), MECS (Medium Energy
Concentrator Spectrometer; 1.3--10 keV) and the PDS (Phoswich Detector System; 13--300
 keV).
Spectral analyses of these observations are reported in Massaro et al. (2004) and
references therein.
Concerning  1998 \sax~ observations of \mrk, started 3 days before the \asca~ pointing, 
Tanihata et al. (2004) reported the spectral analysis of data obtained by the LECS and the
MECS (see also: Maraschi et al. 1999, Fossati et al. 2000a, 2000b, Massaro et al. 2004). 
In May 1999 \mrk~ was pointed for about 4 days; two long uninterrupted observations were 
performed from 26 April and 3 May, 2000, followed by another observation from 9 May 2000
and 12 May 2000.
In all observations of 1999 and 2000, \mrk~ showed a strong  variability on time scales of
 days. Data reduction and spectral analysis of these \sax~ observations are described in
the references above.
\begin{table}
\caption{Spectral parameters  of the LP model of time resolve spectra for the 0.5--10
\xmm~
EPIC-MOS observations of Mrk 421}
\begin{flushleft}
\begin{tabular}{lrrrr}
\hline
\noalign{\smallskip}
$Date$          & $E_p ^*$                     &   $b$          &$S_p^{**}$                     &$\chi^2_{r}~(d.o.f.)$ \\
\hline 
\noalign{\smallskip}
01/11/00        & 0.22~(0.08)                & 0.25~(0.05)     &132.8~(9.6)                  & 0.86~(174)       \\ 
                & 0.22~(0.04)                & 0.23~(0.02)     &124.8~(4.8)                  & 0.95~(307)       \\
                & 0.31~(0.06)                & 0.28~(0.03)     &136.0~(4.8)                  & 1.12~(225)       \\
13/11/00        & 1.09~(0.04)                & 0.27~(0.02)     &340.6~(1.0)                  & 1.06~(372)       \\
                & 0.92~(0.03)                & 0.40~(0.02)     &342.9~(1.1)                  & 1.07~(363)       \\ 
08/05/01        & 0.74~(0.05)                & 0.35~(0.02)     &254.4~(1.6)                  & 1.01~(285)       \\
                & 0.82~(0.04)                & 0.27~(0.02)     &263.5~(11)                   & 1.23~(347)       \\
                & 0.85~(0.04)                & 0.37~(0.02)     &249.4~(1.1)                  & 1.11~(311)       \\
04/05/02        & 0.39~(0.07)                & 0.49~(0.06)     &147.2~(6.4)                  & 0.98~(139)       \\
                & 0.38~(0.06)                & 0.46~(0.05)     &144.0~(6.4)                  & 1.06~(154)       \\      
                & 0.44~(0.06)                & 0.56~(0.06)     &120.0~(4.8)                  & 0.96~(153)       \\      
05/05/02        & 0.23~(0.04)                & 0.44~(0.04)     &102.4~(6.4)                  & 0.98~(186)       \\      
14/11/02        & 1.43~(0.04)                & 0.39~(0.03)     &347.2~(1.6)                  & 1.00~(221)       \\      
                & 2.0 ~(0.1)                 & 0.32~(0.05)     &380.8~(4.8)                  & 1.22~(155)       \\      
01/12/02        & 0.87~(0.05)                & 0.41~(0.03)     &172.6~(1.1)                  & 1.01~(228)       \\      
                & 0.72~(0.07)                & 0.38~(0.04)     &169.6~(1.2)                  & 1.10~(195)       \\      
                & 0.67~(0.06)                & 0.40~(0.04)     &167.7~(1.4)                  & 0.91~(212)       \\      
02/12/02        & 0.64~(0.05)                & 0.34~(0.03)     &196.8~(1.6)                  & 1.00~(216)       \\      
                & 0.83~(0.08)                & 0.35~(0.04)     &227.2~(1.6)                  & 1.08~(190)       \\      
02/06/03        & 0.36~(0.08)                & 0.45~(0.06)     &51.2~(3.2)                   & 1.09~(139)       \\      
14/11/03        & 1.20~(0.03)                & 0.43~(0.02)     &467.2~(1.6)                  & 1.15~(321)       \\      
                & 0.79~(0.04)                & 0.48~(0.03)     &441.6~(3.2)                  & 1.10~(240)       \\      
10/12/03        & 0.87~(0.05)                & 0.38~(0.03)     &209.1~(1.1)                  & 1.04~(274)       \\      
                & 0.89~(0.05)                & 0.42~(0.03)     &207.2~(1.3)                  & 1.17~(240)       \\      
06/05/04        & 1.74~(0.04)                & 0.33~(0.03)     &432.0~(1.6)                  & 1.09~(332)       \\      
                & 2.4~(0.1)                  & 0.24~(0.02)     &494.4~(3.2)                  & 1.24~(335)       \\      
07/11/05        & 0.9~(0.1)                  & 0.38~(0.06)     &284.8~(3.2)                  & 1.06~(145)       \\      
                & 0.8~(0.1)                  & 0.44~(0.07)     &296.0~(4.8)                  & 1.04~(125)       \\      
09/11/05        & 0.63~(0.03)                & 0.40~(0.02)     &420.8~(3.2)                  & 1.27~(365)       \\      
                & 0.71~(0.04)                & 0.41~(0.02)     &430.4~(3.2)                  & 1.25~(302)       \\      
                & 0.68~(0.04)                & 0.36~(0.02)     &460.8~(3.2)                  & 1.16~(300)       \\      
                & 0.79~(0.05)                & 0.40~(0.03)     &505.6~(3.2)                  & 1.22~(266)       \\      
\noalign{\smallskip} 
\hline
\end{tabular}
\end{flushleft}
(*) $E_p$ in keV

(**) $S_p$ in 10$^{-12}$erg cm$^{-2}$ s$^{-1}$ 
\end{table}

\section{Spectral Analysis}
\begin{figure*}[!htp]
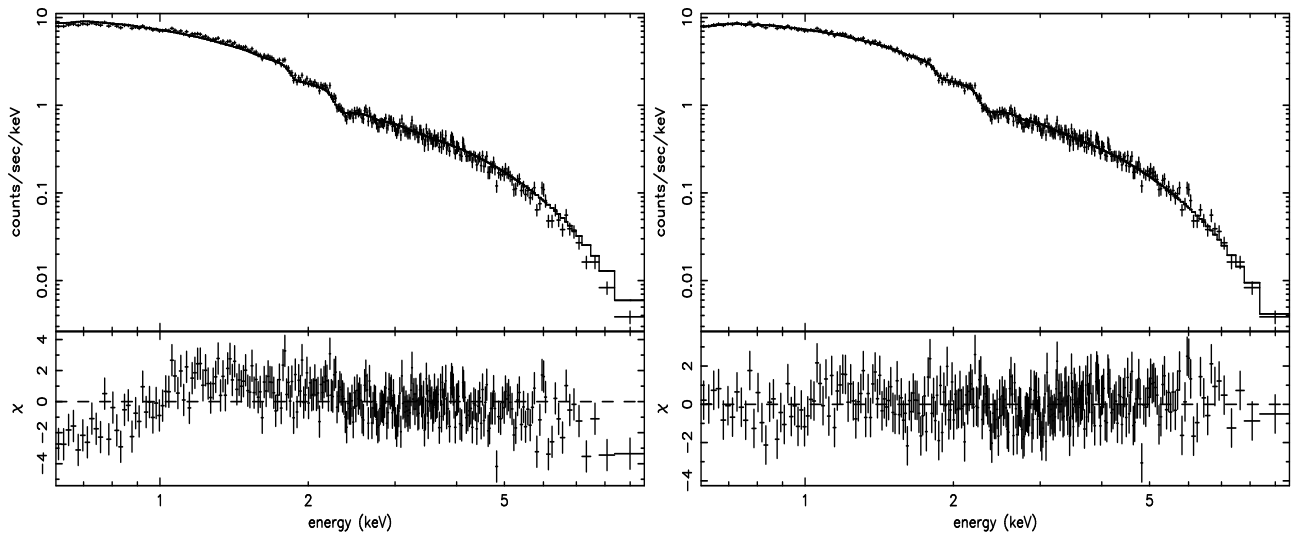

\begin{tabular}{ c c}
\includegraphics[height=8.5cm,width=7cm,angle=-90]{fig1a.ps}
\label{}
\includegraphics[height=8.5cm,width=7cm,angle=-90]{fig1b.ps}
\label{}
\end{tabular}
\caption{\xmm~ EPIC-MOS spectrum from the observation of Mkn~421 performed on 08/05/01.
\textit{Left}: the systematic deviations on both sides of the residuals from a best fit PL
with Galactic \nh show the need of intrinsic curvature. \textit{Right}: the deviations
disappear with the LP model with Galactic \nh.}
\end{figure*}
Our \xmm~ spectral  analysis was  performed with  the {\sc xspec} software
package, version 11.3 (Arnaud, 1996).
Before evaluating  any  spectral curvature in the  0.5--10 keV range, it is important
to reliably assess the absorption at low energies due to interstellar gas.
In the optical, high resolution images of the host, early-type galaxy of \mrk~ do not
show in the brightness profile any evidence of large amounts of absorbing material  (Urry
et al. 2000). In X-rays, it has been shown by Fossati et al. (2000b) and confirmed by
Tanihata et al. (2004) and Massaro et al. (2004) that describing the spectral shape in
terms of absorption not only would require a column density much higher than the Galactic
value $N_H=1.61\times 10^{20}$cm$^{-2}$ (Lockman \& Savage, 1995), but also would yield in
any case unacceptable fits with very high $\chi^2_r$. These findings motivate us to
consider intrinsically curved spectra with the column density fixed at the Galactic value,
in agreement with the above analyses of \mrk. 

In fact, we tried a number of different spectral models:
single (PL) and broken (BPL) power-laws; a power-law with an exponential cutoff at high
energy (PLC); and a logarithmic parabola (LP) in the form

\begin{equation}
F(E) = K~E^{-a-b~\log~E}~, 
\end{equation}
where $a$ is the spectral slope at $E_1 =1$~keV,  and $b$ measures the
curvature.
 
We find that the PL model is not adequate to describe the spectral shape, consistent with 
other X-ray analyses reported above. In our  observations we mostly obtained unacceptable
values for $\chi^2_r$  (higher than 1.5); in the few cases where the $\chi^2_r$ is
acceptable,  the residuals show systematic deviations clearly indicating 
the presence of a curvature, see Fig. 1.
PLC and BPL models give better spectral descriptions, but they also have drawbacks. 
The PLC model often estimates the cut-off energy at much higher than 10 keV, indicating 
that in the instrumental range there is a mild curvature rather than  a sharp exponential
cutoff. The BPL model gives systematic excess at high energies.
The LP model describes the spectral shape more satisfactorily than all the
above  ones, giving  systematically  lower  $\chi^2_r$  values and fewer residuals. 

The  main  goal   of  our  analysis  is  the search  of  possible
correlations between the  spectral curvature ($b$),  the SED peak energy ($E_p$), and the 
corresponding  SED peak value ($S_p$). 
To estimate the correlation between the parameters of the spectral energy distribution
$S(E)=E^{2}F(E)$ we often  use in place of Eq. (1) the equivalent functional form (LPS),
 see Sect. 3 in Massaro et al. 2004 and Tanihata et al. 2004
\begin{equation}
S(E) = S_p~10^{-b~\log^2(E/E_{p})}~, 
\end{equation}
where  $S_p=E_{p}^{2}\, F(E_p)$. In this form the values of the parameters $b$, $E_p$ and 
$S_p$ are estimated  in the fitting procedure, whereas those are derived from Eq. (1) are
affected by intrinsic correlations. For our analyses,  we added  a specific spectral
modelling routine to the standard  {\sc xspec} v. 11.3 software package.

We shall see that the estimates of the curvature parameter $b$ are  sensitive to changes 
of the source state during a pointing; so a time averaged spectrum may have a different
curvature from the the time resolved ones.
To avoid any such bias, we segmented our data at times when significant (3 sigma) count 
rate variations occur, subject to the constraint of having in each time resolved spectrum
more than 120 spectral bins after rebinning  as indicated in \S 2.      
The duration of our time segments was shorter than a few 10$^4$ s.

An example of this effect is apparent from the two light curves plotted in Fig. 2
in the energy range 0.5-10 keV (left panel) and 4-10 keV (right panel) of the same 
observation on 13/11/00. 
The significant decrease in the count rate of the higher-energy curve is only barely 
detecable in the total light curve. Then the spectral analysis of the two time regions (A
and B, see Fig. 2) shows a spectral dynamics not apparent 
when the whole data set is considered.
\begin{figure*}[!htp]
\begin{center}
\begin{tabular}{cc}
\includegraphics[height=8.5cm,width=7.5cm,angle=-90]{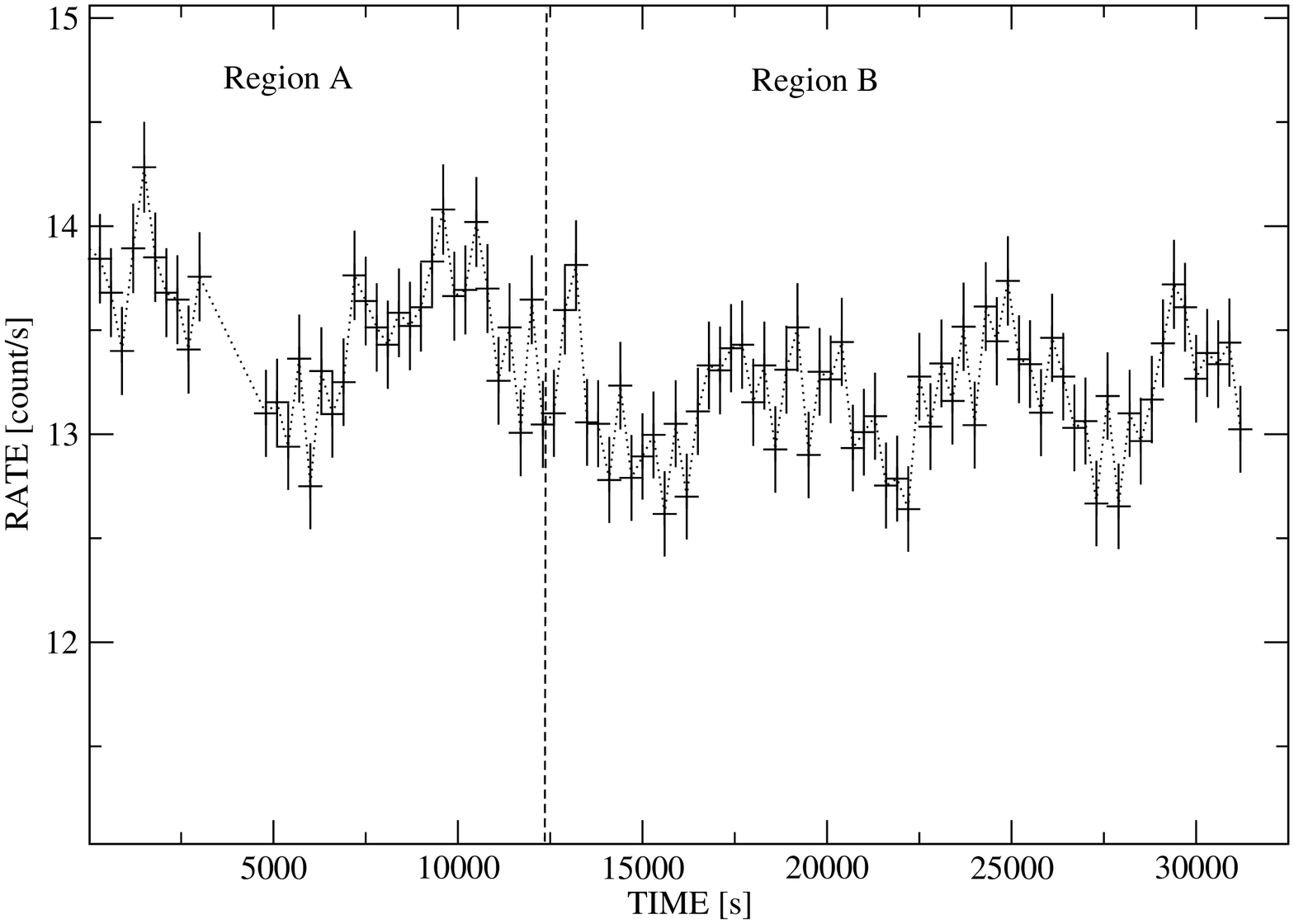}
\label{fig2}
\includegraphics[height=8.5cm,width=7.5cm,angle=-90]{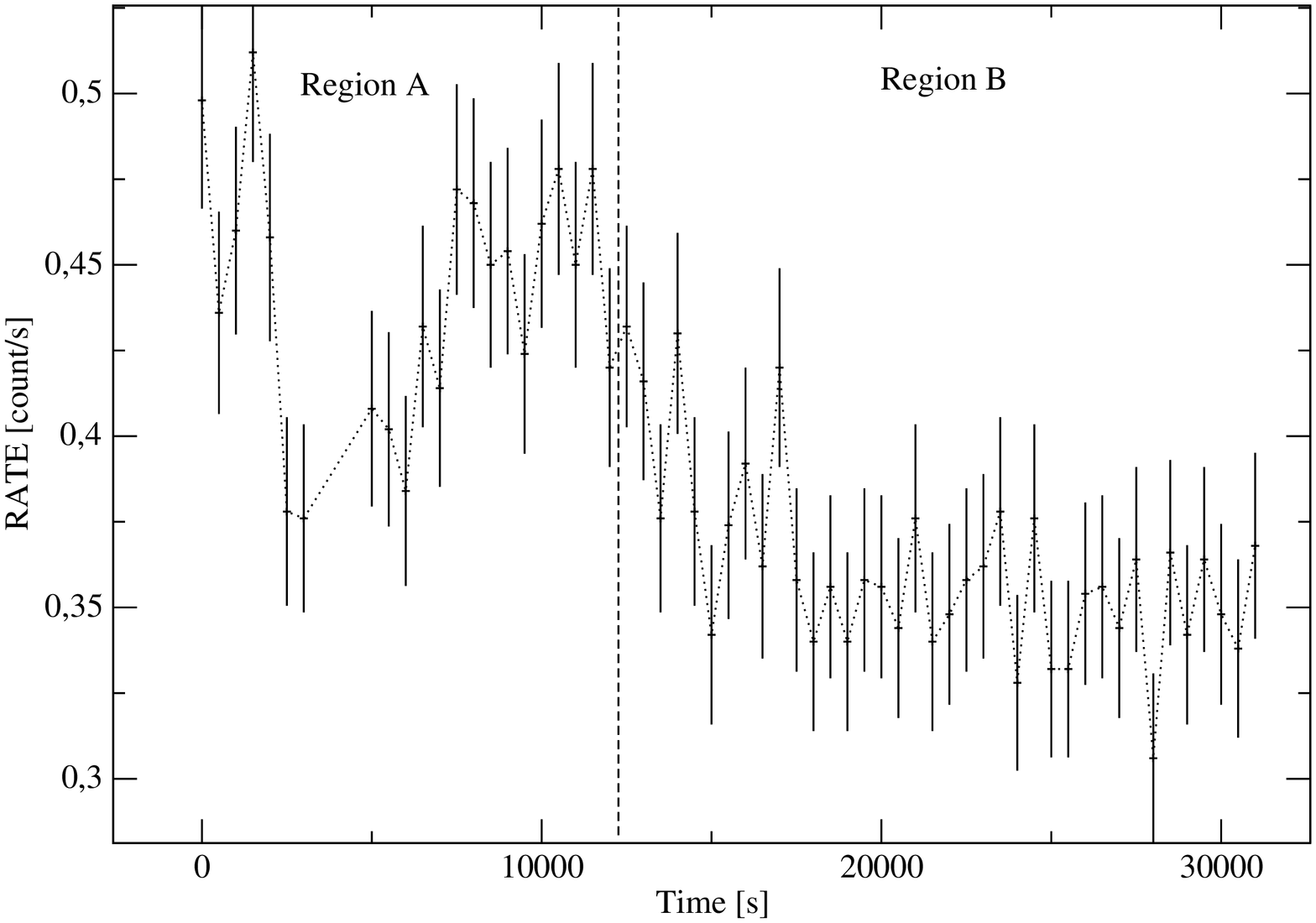}
\label{fig3}
\end{tabular}
\caption{(\textit{Left panel}): \xmm~ light curve in the energy range 0.5-10 keV.
(\textit{Right panel}): \xmm~ light curve in the energy range 4-10 keV, where count rate
variations are evident   between the two selected time regions.}
\end{center}
\end{figure*}

The spectral best fit values for all \xmm~ observations are reported in Table 2.
All statistical errors refer to the 68\%  confidence level (one Gaussian standard 
deviation). 

Our results generally agree with those from observations with \sax, \asca~ and \swf~ 
(see the references above and Tramacere et al. 2006 for the \swf~ analysis).
In particular, we find values of SED peak energy and of the curvature parameter consistent
with their values,  but we find differences with other \xmm~ analyses.
Specifically, Sembay et al. (2002) analysed \xmm~ observations during 2000, 2001 and 2002,
 but they used an energy range restricted to 2-10 keV and did not search for any curved 
shape. Ravasio et al. (2004), who studied EPIC-PN observations performed on 04/11/02,
14/11/02 and 01/12/02,  didn't found a significant improvement using the log-parabolic
model respect to the power-law one. The difference between their and our analysis
is likely  due to our more conservative criterion to subtract solar flares in the data
reduction. 

\section{Testing the dynamics of synchrotron emission}
Correlations between $S_p$ and $E_p$ provide interesting indications concerning
the driver of the spectral changes in  X-rays, in terms of the synchrotron emission
mechanism from one dominant homogeneous component. In this
framework (Rybicki \& Lightman, 1979) the dependence of $S_p$ on $E_p$ is in the form of a
power-law: 
\begin{equation}
  S_p \propto E_p^\alpha ~.
\end{equation}

In fact, the synchrotron SED is expected to scale as $S \propto N~\gamma^{3}~B^2~\delta^4$
at the energies  $E \propto \gamma^2B~\delta$, in terms of total emitter number $N$, the
magnetic field $B$,  the typical electron energy $\gamma mc^2$, and the beaming factor
$\delta$,
Thus $\alpha=1.5$ applies when the spectral changes are dominated by variations of
the electron average energy;
$\alpha=2$ as for  changes of the magnetic field;
$\alpha=4$  if  changes in  the beaming factor dominate;
formally, $\alpha=\infty$, applies for changes only in the number of emitting particles.

On a heuristic stand, inspection of Fig. 3 indicates that the two last cases are 
unlikely; next we proceed to a formal statistical analysis.
\subsection{Correlation between $S_p$ and $E_p$}
We investigate whether and how  $S_p$ and $E_p$ correlate.
The analysis is performed on two data sets: the \X set and the \FULL set .
The former corresponds to the results of the spectral analysis in \S 3, while the
latter is obtained on adding the results from previous analyses as reported in Sect. 3.\\

We test for correlations between the spectral parameters  on using both a
linear correlation coefficient ($r_{lin}$) and a logarithmic one
($r_{log}$), and fit the data with a linear and log-linear model
following the statistical methods described in D'Agostini (2005) (see also Appendix
A).

Results from our correlation study are given in Table 3. 
The correlation coefficient $r_{log}\simeq0.7$ is significant, specifically with a
chance probability value close to zero. The data in the \X set may be described by a
power-law function as in Eq. (3);
the value of $\alpha$ so obtained  is $\alpha$ $=$ $1.4\pm0.3$ . On fitting the \FULL
set we find the power-law index $\alpha = 1.2\pm0.1$, consistent with the
above value. Even if we took the range of $E_p$ between $0.5-1.0$ keV where the
run of the \FULL data apparently steepens, the values of the best fit would yield 
$\alpha = 1.2 \pm0.1$, quite far from 4. This analysis confirms that the cases
$\alpha=1.5$ and $\alpha=2$ are those most relevant as dominant mechanisms.\\

A closer analysis and an interpretation of these results are given in the next subsection.

\begin{figure*}[!htp]
\begin{center}
\includegraphics[width=10cm,angle=-90]{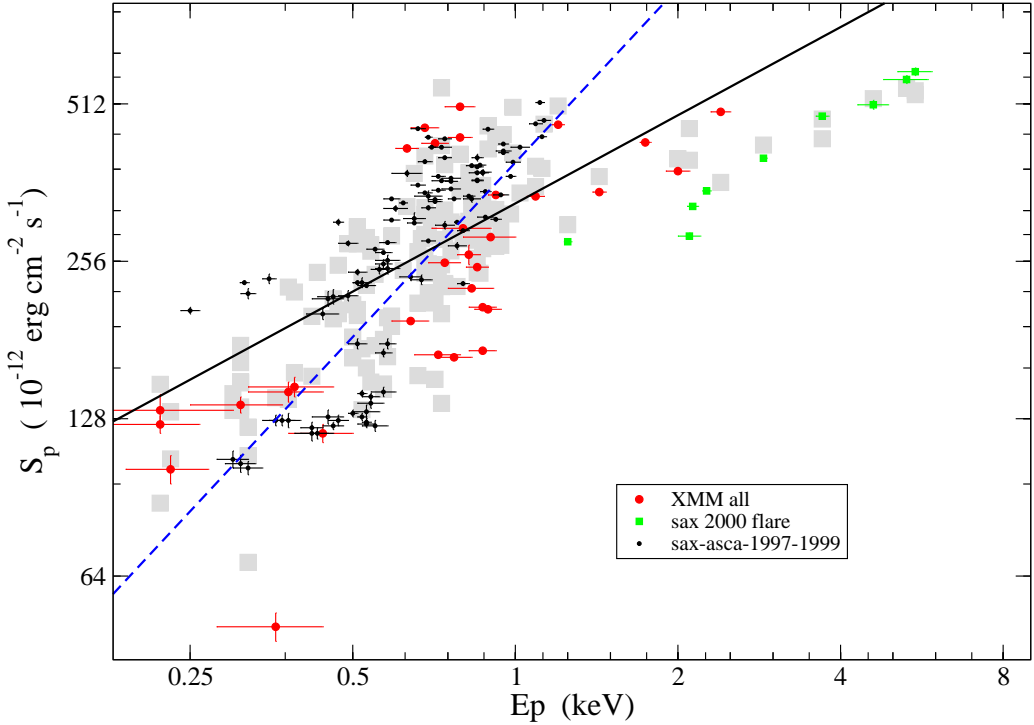}
\label{fig3}
\caption{The  scatter plot  of the peak energy $E_p$ and the maximum of
the SEDs $S_p$ compared with
 the results of a Monte Carlo simulation (grey squares). Dashed line represents the
power-law best fit without taking into
 account extravariance for the \FULL data-set. Solid line represents the power-law best
fit taking into account
 extravariance for the \FULL data-set.}
\end{center}
\end{figure*}

\begin{table*}[!htp]
\caption{Statistical parameters for the correlation between $E_p$ and $S_p$.}
\begin{flushleft}
\begin{tabular}{llllllll}
$Data set$  &$r_{lin}$ &$p_{lin}$ &$r_{log}$   &$p_{log}$ &$\alpha$
&$\alpha_v$ &${\sigma_v}$  \\
\hline 
\noalign{\smallskip}
\X                     &$0.63$   &$<0.001$  &$0.71$ &$<0.001$     &$1.4\pm0.3$ &$0.7\pm0.1$   &$0.35\pm0.05$\\
\FULL                  &$0.56$   &$<0.001$  &$0.67$ &$<0.001$     &$1.2\pm0.1$ &$0.56\pm0.05$ &$0.33\pm0.02$\\
\FULL($E_p<1.2$)  &$0.72$   &$<0.001$  &$0.72$ &$<0.001$     &$1.7\pm0.1$ &$0.99\pm0.07$ &$0.29\pm0.02$\\
\FULL($E_p>1.2$)  &$0.87$   &$0.004$   &$0.84$ &$0.007$      &$0.6\pm0.1$  &$0.43\pm0.08$ &$0.13\pm0.03$\\
\noalign{\smallskip} 
\hline 
\end{tabular}
\end{flushleft}
\end{table*}
\subsection{Signatures of the emission process} 
 A sensitive point in our statistical analysis and its interpretation is
the presence of ``hidden'' parameters contributing to variability and/or to correlations.
On the basis of the formalism that for reader's convenience we recall in Appendix A, in
the following we discuss in details the possibility that a correlation may be introduced
by a parameter that is not directly determined from the
data (whence the name "hidden").

This issue arises when one estimates two parameters, say $x$ and $y$, 
statistically independent but depending physically on a third parameter, $k$ say, that is
not directly observed (``hidden''); in other words, we observe $x=x(x', k)$ and
$y=y(y', k)$. Again, a correlation will arise between $x$ and $y$ due to their dependence 
on $k$, and this may add to  any physical correlation between $x'$ and $y'$.

A case in point is provided by the beaming effects.
In BL Lacs  beaming is a key property; though only rarely  measured directly, this
can introduce dependence between the observed quantities $S_p$ and $E_p$. It is convenient
to distinguish the variables $E_p'$, $S'_p$ in the beam reference frame from the observed
ones expressed as 
\begin{eqnarray}
E_p&=&  E_p'\delta \\ \nonumber 
S_p&=& S'_p\delta^4.
\end{eqnarray}\\
While this corresponds to the case $\alpha=4$ that our preliminary analysis tended to
exclude, it is important to reconsider   $\delta$ as a hidden parameter  introducing a
covariance term and contributing to correlation.\\

Clearly, the actual  contribution depends on the probability density functions (PDF) of 
$E_p$, $S_p$ and $\delta$. To study this, we have generated with a Monte Carlo code a
sample of uncorrelated pairs of variables $(x',y')$. We then have transformed them
according  to Eq. (4), that is, with $x=\delta x'$ and $y=\delta^4 y'$, choosing
$\delta=10$ as a typical value for \mrk.
The $x'$ variables were generated so that the $x$ variables have the same PDF as $E_p$;
on the other hand, the $y'$ variables were generated from a normal distribution with mean 
value $\mu'=1.0$ and $\sigma'=\mu'/3$.
We impose that the $y$ variables have the same dispersion as  $S_p$ in the \FULL
data;
this dispersion depends not only on $\sigma'$ but also on the distribution of
$\delta$, so we have to derive a constraint on the value of  $\sigma_{\delta}$.
In fact, the standard deviation of the observed values of
$S_p$ is about 120 (erg~$\rm ~cm^{-2}$~ $\rm s^{-1}$) to be compared with the average
value of about 290 (Fig. 3); assuming, as an extreme case,  that the dispersion
$\sigma_{S}$ of $S_p$ is generated only by the variance $\sigma_{\delta}$, and applying
standard propagation, we obtain  $\sigma_{\delta}/\mu_{\delta}=\sigma_{S}/(4~S_p) 
\lesssim 10\% $. 

We now generate  $\delta$ from a normal distribution with $\mu_{\delta}=10$ and
$\sigma_{\delta}$ = $0.75$, and find a correlation coefficient  $r_{log}=0.3$ between the
logarithms of $x$ and $y$.
On decreasing the variance of $y'$ and increasing that of $\delta$,  the correlation
increases but only slightly; for example, if we take $\sigma'= \mu'/ 6$ and increase 
the value of $\sigma_{\delta}$ to $0.95$, we obtain $r_{log}=0.36$.

In summary, the beaming factor can affect the observed correlation coefficient, but only
up to values
$r_{log} < 0.3$ which are significantly below the observed value $0.67$.
We conclude that  beaming alone is responsible neither for the values of $\alpha \approx
1$  nor for the correlation observed.
We stress that the value $\sigma_{\delta}/\delta \lesssim 0.1$, bounded by the dispersion
of the full data,  is remarkably low when compared to the average  typical value of
$\mu_{\delta} \simeq 10$, and implies the beaming factor to have been closely constant for
\mrk~ during our observational span of about 9 yr. 

It follows that both the variations of $S_p$ corresponding to the estimated values of 
$\alpha$ and their scatter must be importantly contributed by a physical
process in the beam rest frame, such as variations of magnetic fields  or scaling up or
down of all the electron energies.

In fact, a second effect coming from a ``hidden" parameter with its fluctuations is  to  
 produce   scatter of the data. As explained in Appendix A, it is
possible to account for this effect  by adding an ``extravariance" in the 
likelihood function.
In the case of the \X data set, the extravariance accounting for their scatter 
is  estimated at $\sigma_{v}=0.35\pm0.05$, and for the corresponding value of the slope we
obtain $\alpha_v=0.7\pm0.1$, whilst for the \FULL data set we obtain
$\sigma_{v}=0.33\pm0.02$ and $\alpha_v=0.56\pm0.05$ (see Table 3).
 The derived values of the $\alpha$ index are considerably affected by the
extravariance; in fact,  the extravariance term in the log-likelihood function dominates
the term from measurement uncertainties,  to the point of providing power-law indices 
close to those obtained simply from fitting the data with no  weights for their
precision.

Such  different values of $\alpha$ and $\alpha_{\nu}$ (Tab. 3) lead us to test an
actual change in the power-law index in Eq. (3) at about 1 keV.
To do this,  we split the \FULL set into two sets, one with $E_p <1.2$ keV and the other
with $E_p > 1.2$ keV; results of fits are reported in the last two lines of Table 3.
The values of $\alpha$ found for the two data sets differ significantly, being
$\alpha_1=1.7\pm0.1$ and $\alpha_2=0.6\pm0.1$, for  $E_p<1.2$ keV and $E_p>1.2$ keV, 
respectively.
This  could be interpreted as a change of $\alpha$ with the state of the source, and we
estimate  the associated likelihood  by means of a Monte Carlo simulation.
We generate  a set of values for $E_p$, $S_p$ as follows:
\begin{itemize}

\item $E_p$ values are generated as to have the same PDF as observed once
transformed by the first of Eq. (4).

\item Beaming factors are generated from a normal distribution
with $\mu_{\delta}=10$ and $\sigma_{\delta}=0.5$ .

\item The slope of the (unbeamed) power-law  is generated  from two  normal distributions 
with mean value $0.85$ and standard deviation $0.09$ for $E_p<1.2$ keV, and mean value
$0.33$ and standard deviation $0.07$ for $E_p>1.2$ keV.
 
\item Values of  $S_p$  are generated from Eq. (3) and transformed by the second of Eq.
(4).
\end{itemize}
The number of events so generated is $150$ (similar to the \FULL data set), and in Fig. 2 
we report their scatter plot (grey box). Interestingly, the simulated and the observed
points have similar behaviour and similar statistical properties. 
We see that the correlations, given in the beam reference frame by $\alpha_1$ and
$\alpha_2$, when subject to beaming with a narrow variance come very close to account for 
the data with their scatter.

\section{Probing signatures of the electron acceleration}
Finally, we perform  a study of correlation between $b$ and $E_p$, aimed at pinpointing   
possible signatures of the electron acceleration processes in the spectral
evolution  of a homogeneous dominant component.
The same data set and statistical tools of the previous section are used. 


\subsection{Test of correlation between $b$ and $E_p$}
Such a correlation had never been tested previously. 
An analysis of \X set shows that the correlation coefficients are
negative and low, that is, $r_{lin}=-0.29$  and $r_{log}=-0.13$ (defined in Sect. 4).
The former corresponds to a low chance occurrence ($0.05$), but this is significantly high
($0.24$) for  $r_{log}$.
By direct  inspection we see that there are three points (enclosed by the dashed ellipse 
in Fig. 4) corresponding to the same \X pointing on 01/11/00, that  maximally 
deviate  from the average sample behaviour. 
On excising this observation and reanalysing the remaining data (denoted by \X* in Table
5)  we obtain the substantially higher correlation coefficients $r_{lin}=-0.60$ and
$r_{log}=-0.62$, corresponding to the high significance, given in  Table 4.

To see whether the 01/11/00 observation constitutes a unique event, we analysed the \FULL 
data set with and without (denoted by \FULL*) this pointing; Table 4 shows that the values
of $r_{lin}$ and $r_{log}$ for \FULL and \FULL* are close, indicating that the observation
excised has not materially changed the overall statistical behaviour. 

These values are sufficiently
high to warrant a brief discussion of the importance of the correlation found, along two 
ways. This pointing may constitute just a rare source state  in our sample.
On the other hand, this  may not constitute a singular event,  considering that
all \X and \FULL data (Fig. 4) appear to outline an upper limit to the curvature observed 
at any  values of $E_p$. For example,  for $E_p$ around 1 keV, we never observe curvatures
higher than about $b \approx 0.45$.
\begin{figure*}[!htp]
\begin{center}
\includegraphics[width=10.0cm,angle=-90]{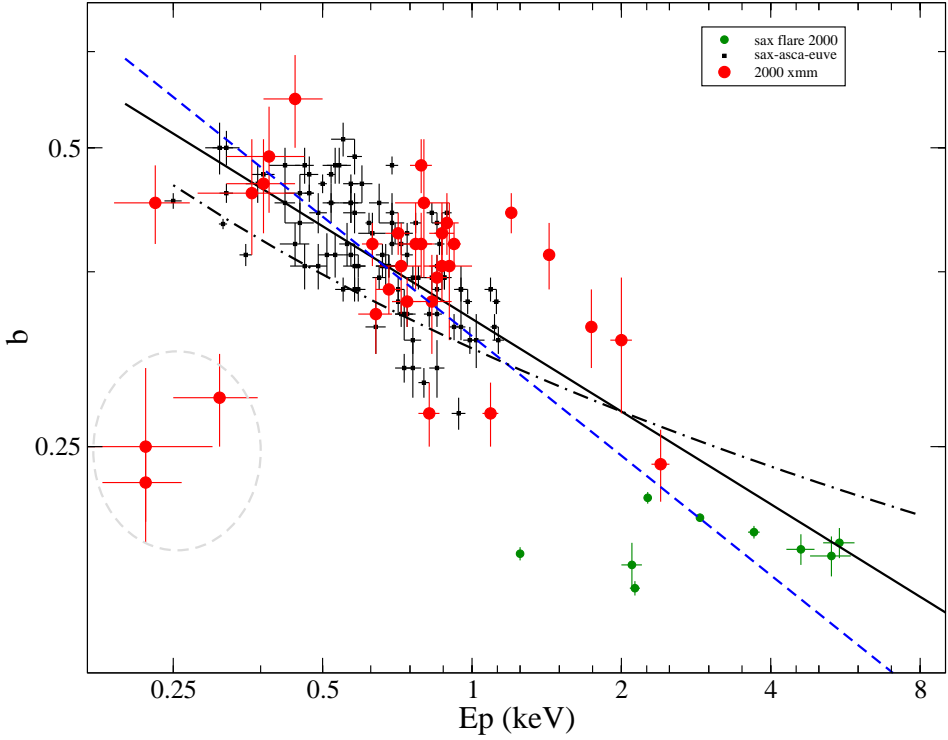}
\label{fig5} 
\caption{The  scatter plot of the peak energy $E_p$ and the curvature parameter $b$.
  The dashed line represents the power-law best fit without taking into account
  extravariance for the \FULL data-set. The solid line represents the power-law best fit 
on taking into account extravariance for the \FULL data-set. The dashed-dotted
  line represents the relation between $E_p$ and $b$ expected from Eq. (11)}
\end{center}
\end{figure*}
\begin{table*}[!htp]
\caption{Statistical parameters for the correlation between $E_p$ and $b$}
\begin{flushleft}
\begin{tabular}{llllllll}
$Data set$   & $r_{lin}$ &$p_{lin}$ &   $r_{log}$ &$p_{log}$ &$\alpha_{pl}$
&$\alpha_{\sigma_v}$ &${\sigma_v}$    \\
\hline
\noalign{\smallskip}
\X           &$-0.29$    &$0.05$   &$-0.13$           &$0.24$   &$-0.04\pm0.03$   &$-0.04\pm0.06$  &$0.18\pm0.03$         \\
\X*          &$-0.60$    &$<0.001$ &$-0.60$           &$<0.001$ &$-0.20\pm0.03$   &$-0.22\pm0.06$  &$0.12\pm0.02$         \\
\FULL        &$-0.63$    &$<0.001$ &$-0.70$           &$<0.001$ &$-0.404\pm0.005$ &$-0.31\pm0.02$  &$0.13\pm0.01$         \\
\FULL*       &$-0.67$    &$<0.001$ &$-0.79$           &$<0.001$ &$-0.404\pm0.005$ &$-0.34\pm0.02$  &$0.12\pm0.02$         \\
\noalign{\smallskip} 
\hline
\end{tabular}
\end{flushleft}
\end{table*}
\subsection{Interpretations}
The inverse correlation between  $b$ and $E_p$ may be interpreted in the framework of
acceleration processes of the emitting electrons. \\
A first interpretation of the correlation between $b$ and $E_p$ is in the
framework of statistical acceleration (Massaro et al. 2006 and references therein). In 
this scenario  the probability for a particle to be further accelerated decreases at high
energies, being inversely proportional to the energy itself. For example, this may occur
when the particles are confined by a magnetic field, and the confinement efficiency
decreases as the gyration radius increases.

In such cases the electron energy distribution is curved into a log-parabolic shape, and 
its curvature $r$ is related to the fractional acceleration gain $\epsilon$ as  given by
Massaro et. al. (2006): 
\begin{equation}
 r\propto \frac{1}{\log \, \epsilon}~.
\end{equation} 
Note that $r$ decreases  when $\epsilon$ increases.
The spectrum of the synchrotron emission from these particles is also curved, with 
\begin{equation}
 b \approx \frac{r}{5}~.
\end{equation}
On the other hand, $E_p$ scales like $\epsilon$, so a negative
correlation between $b$ and $E_p$ is expected. This basically arises from a loss of
acceleration efficiency  at high energies.

On the other hand, a connection of log-parabolic spectra with acceleration may be also
understood in the framework provided  by the Fokker-Planck equation (Kardashev 1962)

\begin{equation}
  \frac{\partial N}{\partial t} =    \frac{\partial}{\partial \gamma}
       \Big(\lambda_1 \gamma^2 \frac{\partial N}{\partial \gamma}\Big)  
     -\frac{\partial} {\partial \gamma}\Big(\lambda_2 \gamma N\Big)  
     ~.
\end{equation}
This describes  the evolution of the distribution function $N(\gamma,t)$ of the electron 
energies $\gamma\, m\,c^2$ 
under stochastic (first term on the r.h.s.) and systematic acceleration (second term on
r.h.s.). 
The simple  solution given by Kardashev (1962) for an initial, monoenergetic 
injection at the energy $\gamma_0\, m c^2$ reads 
\begin{equation}
  N(\gamma,t) \propto \rm \gamma^{-1}~exp\Big[\frac{-(\Lambda_1 +
  \Lambda_2-ln(\gamma/\gamma_0))^2}{4\Lambda_1}\Big]~, 
\end{equation}
with

\begin{eqnarray}
\Lambda_i = \int_{0}^{t} \lambda_i(t)~dt  \nonumber.
\end{eqnarray}\\
Eq. (6) again describes a log-parabolic distribution, with the curvature term

\begin{equation}
r=\frac{1}{4\Lambda_1},
\end{equation}
inversely proportional to the coefficient of the random acceleration component.

On the other hand, the peak energy of $N(\gamma,t)$ is given by
 \begin{equation}
  \gamma_{max} = \gamma_0~e^{\Lambda_2 - \Lambda_1}    \gamma_0 e^{ \Lambda_2  - 1/4 r}.	
\end{equation}
The  logarithm of the peak energy $E_p$ of the SED is closely proportional to the
position of the peak of the  logarithm of $\gamma^3\, N(\gamma)$. If  $N(\gamma)$ has
 a log-parabolic shape, using again Eq. (6) that applies to  any log-parabolic energy
distribution, we obtain
\begin{equation}
\ln \, E_p \propto 2\ln\,\gamma_{max} + \frac{3}{5 b}.\\
\end{equation}
In terms of $S_p$ and $E_p$ we see  the logarithm of $E_p$ to be inversely proportional
to $b$; this is consistent with the inverse  correlation observed between $b$ and $E_p$.

In this framework the inverse correlation constitutes the signature of a stochastic 
acceleration process that  broadens $N(\gamma,t)$
as to decrease its curvature while  $E_p$ is driven to higher values.
\begin{figure*}[htp]
\begin{center}
\begin{tabular}{l l l}
\includegraphics[width=7.5cm]{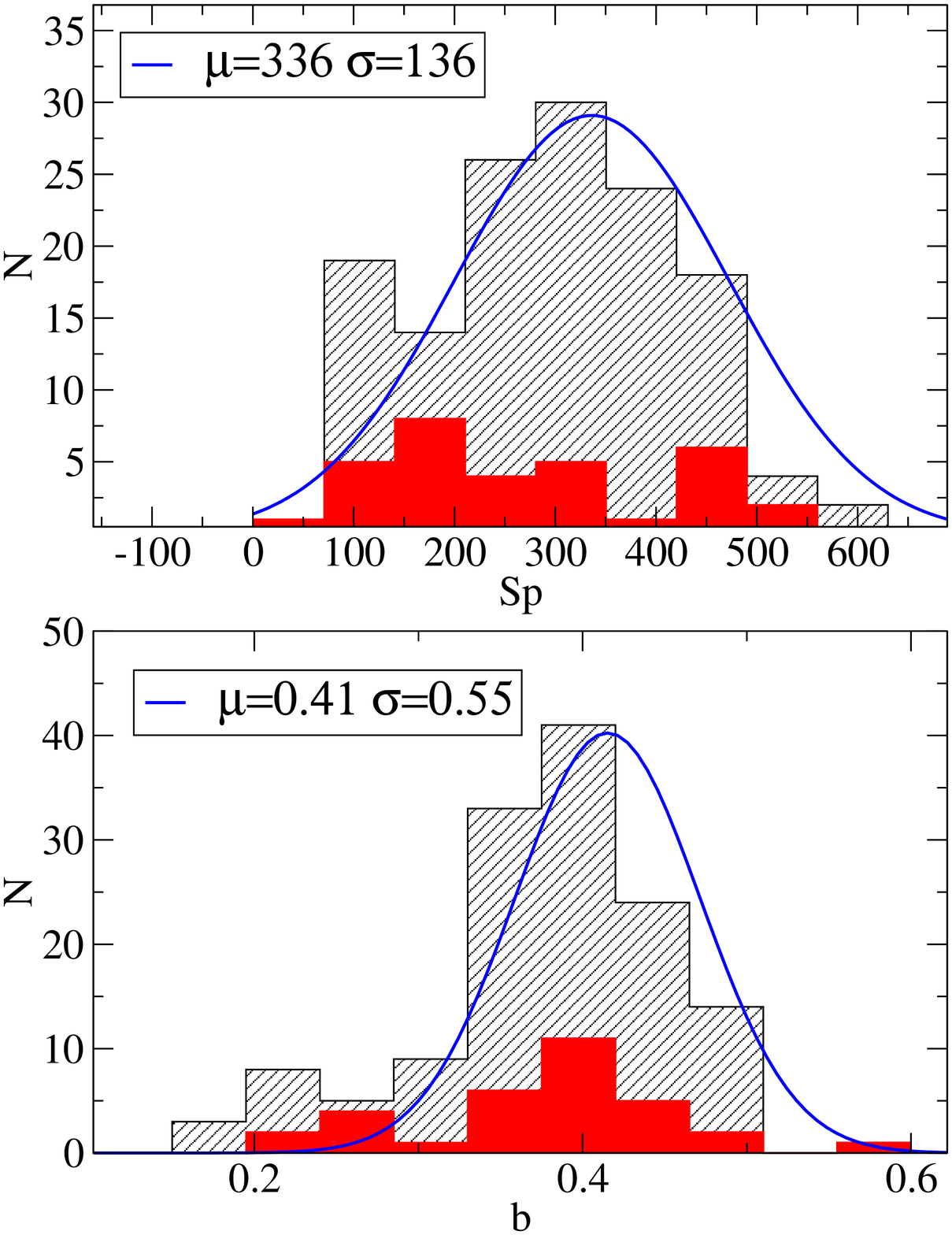}
\label{}
\includegraphics[width=7.5cm]{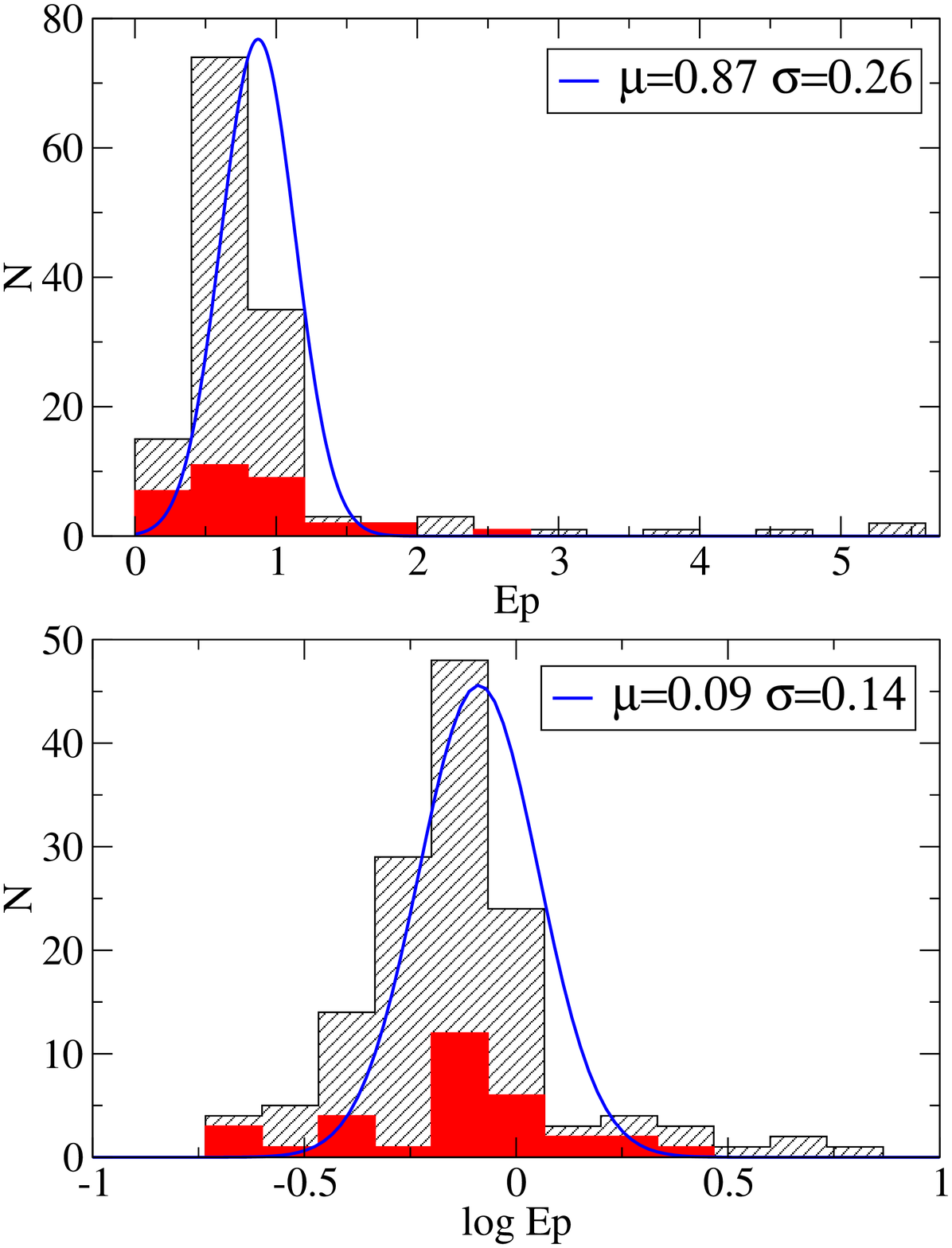}
\end{tabular}
  \caption{Histograms of $S_p$, $b$ and $E_p$: filled boxes represent the \X data set,
 while hatched boxes represent the \FULL data set. The solid lines show
 best fits with a Gaussian, with average ($\mu$) and standard deviation ($\sigma$) given
 in the labels. }
\end{center}
\end{figure*}
\section{Conclusions and discussion}
Using a large set of X-ray observations of \mrk, 
we have presented and discussed the results concerning the spectral variations. We have
shown that
correlations  exist between the peak values $S_p$ of its SED and the peak positions $E_p$,
and between the spectral curvature $b$ and  $E_p$.

The former may be interpreted in the framework of synchrotron emission. The values of the 
power-law slopes obtained from our fits in Sect. 4 are  bounded by $\alpha \leq 1.2 \pm
0.1$. As such, they  rule out the case  $\alpha = 4$ applying  if the beaming factor
$\delta$ were the dominant driver of the spectral evolution; they are instead
compatible with a combined  effect of variations of $B$ (corresponding to $\alpha =
2$) and of a rescaling of $\gamma$ ($\alpha= 1.5$). 

In parallel, a secondary role for  $\delta $ is confirmed by the bound $r < 0.3$ it can 
contribute to the  observed  $S_p, \, E_p$ correlation coefficient  value $r \approx
0.7$. On the other hand, we have set an upper limit to the beaming variance; from the
analysis in sect. 4.2 we have found a low fractional variation
$\sigma_{\delta}/\mu_{\delta}\lesssim10\%$ even on considering conservative values 
$\delta \approx 10$. The remarkable implication is that the beaming factor of Mrk 421 
remained closely  constant during a time span of about 9 years.

This limit is relevant in the framework of the internal shock scenario. 
This  assumes that  shells ejected from the central engine with slightly different
relativistic velocities and slightly differing angles collide in the jet at sub-parsec
scales and  produce flares. The temporal behaviour  and the radiative efficiency of this
process  depend on the collision frequency and on the collision energetics, respectively; 
two versions are found in the literature.
Guetta et al. (2004) assume that shells are ejected at a frequency 
close to $10^{-4}$~Hz,  with $\Gamma$ values distributed around the average value of about
 $~15$
after a random (uniform) distribution  with a considerable dispersion, about 3.
 The dispersion is considerably larger than the values obtained from our
analysis.
On the other hand, Tanihata et al. (2003) assume values of $\delta$ following a
normal distribution with $\sigma_{\delta}/\mu_{\delta}\ll 0.1$ and ejection intervals
around 600 s; whence they obtain a good reproduction of the temporal behaviour, but also a
 very low radiative efficiency.
The upper limit derived from our analysis, much lower than the value assumed by 
Guetta et al. (2004),  emphasizes the efficiency problem reported by 
Tanihata et al. (2003). 

The correlation we have observed in Mrk 421 between  $b$ and $E_p$  is interesting  in the 
framework of the electron acceleration mechanisms. In \mrk~ 
we systematically observed a decrease of the curvature $b$ as the peak energy $E_p$ 
increased. To understand this behaviour we have used in Sect. 5 a  Fokker-Planck
description of  a dominant electron energy distribution.  The solution of this
equation for an initial mono-energetic injection  predicts  (see Eqs. 8-11)  
that with ongoing stochastic  acceleration  the curvature should decrease while 
the peak energy moves to  higher energies. 
A more detailed understanding of this dynamics requires a full  computation including 
radiative cooling and fixing the relative weights of the systematic vs. 
the stochastic acceleration component;  this will be studied
in a different paper (Tramacere et al. 2007 in prep.).
An alternative explanation of this correlation is discussed in Sect. 5 in terms of  
statistical, energy-dependent acceleration probability. This  leads again to  a
correlation as observed.  

We wish to stress a common point to both views, that is, the relevance of the curvature  
parameter to understand the observed spectral evolution of the source. The negative 
correlation between $b$ and $E_p$ strongly indicates  the dynamics of $E_p$ to be related
to stochastic or to statistical (and energy dependent) acceleration mechanisms; it would
not hold with the beaming as the main driver of spectral variations. So the results from
this second correlation  are consistent  with those  from the first, namely, the  $S_p$ --
$E_p$ correlation. 

No significant correlation has been found between $S_p$ and $b$. This lack may arise
from the opposite signs of the correlations between $S_p$-$E_p$ and $E_p$-$b$ adding to
 the considerable dispersion of the data.

Finally, another view on this matter is  provided by the analysis  given in Sect. 5  (see
Fig. 5) concerning the  PDF of the spectral parameters $S_p$,  $b$ and $E_p$.
It is seen that $S_p$ and $b$ have enough symmetry in their PDF to be reasonably
approximated by a Gaussian distribution with minor  deviations in the tail (left panels).
The parameter $E_p$,  on the contrary, shows a more skewed distribution (right upper
panel), that could be better approximated by a log-normal shape, i.e., by a Gaussian in
the variable $\log E_p$ (right bottom panel). We note that a similar distribution has been
successfully used  also to describe the statistical properties of the GRB peak energy
distribution that may depend on broadly similar physics (Ioka \& Nakamura 2002).

A point to stress is that the log-normal distribution constitutes the asymptotic limit 
from the central limit theorem in multiplicative form; in fact, it has been shown by   
Ioka \& Nakamura (2002) that  the limiting log-normal form is closely attained already
after  3 steps.
Stochastic acceleration, for example,  may be treated in terms of  multiplication of  a 
number of random fractional energy gains $\epsilon$, see Eq. (5). The issue will be dealt
with in more detail in Tramacere et al. (2007 in prep.);  here it provides complementary
support to our stress on the relevance of stochastic acceleration to understand the
spectral variations of \mrk. 

\section{Appendix A: Correlations and fits in the presence of uncertainties on both axes
and of hidden parameters}

We first recall the general formalism used to evaluate correlation coefficients for
data with uncertainties on both axes. \\

For two statistical variables defined as: 
\begin{eqnarray}
\bar{\mathbf x}=\{\bar{x_i}\}&=&\{x_i+d x_i\}\\
\bar{\mathbf y}=\{\bar{y_i}\}&=&\{y_i+d y_i\} ~,
\end{eqnarray}
the covariance is expressed  by
\begin{equation}
\rm cov(\bar{\mathbf x},\bar{\mathbf y})=\rm cov(\mathbf x,\mathbf y)+\rm
cov(\mathbf x,\mathbf{d y})+\rm cov(\mathbf y,\mathbf{d y})+\rm 
cov(\mathbf{d x},\mathbf{d y})~, 
\end{equation}
and the correlation coefficient reads 
\begin{equation}
r_{\bar{\mathbf x},\bar{\mathbf
y}}=\frac{\rm cov(\bar{\mathbf x},\bar{\mathbf
y})}{\sigma_{\bar{\mathbf x}}\sigma_{\bar{\mathbf y}}}.
\end{equation}
Here 
$\sigma_{\bar{\mathbf x}}$ is given by
\begin{equation}
\sigma_{\bar{\mathbf x}}=\rm cov(\mathbf x,\mathbf x)+2\rm cov(\mathbf x,\mathbf{
d x})+\rm cov(\mathbf{ d x},\mathbf{d x}),
\end{equation}
and similarly  for $\sigma_{\bar{y}}$.

An issue arising in the analysis of correlations (as anticipated in \S 4 of the main
text), is that ``hidden" parameters -- adding to the ''primary'' ones focused by the
analysis -- can introduce a covariance term, i.e., a correlation. To evaluate its weight
we recall the general expression of the covariance term for functions of random
variables. When we have $M$ functions $(f_1,...,f_M)$ of $N$ random
variables  $x_1...x_N$, the covariance of $f_k,f_l$ is given by Barlow (1989) to read
\begin{equation}
\rm{cov}(f_k,f_l)=\sum_{i}\sum_{j}\Big(\frac{\partial
f_k}{\partial x_i}\, \Big)~\Big(\frac{\partial f_l}{\partial x_j} \Big)\rm\;
{cov}(x_i,x_j)~.
\end{equation}
Eq. (17) shows that even when all the variables $x_i$ are mutually uncorrelated (with
covariance=$0$) a finite covariance term arises for the functions $f_i$ which share the
same variables.\\

We then show explicitly the formalism we actually use in fitting our  data. 
We follow the approach of D'Agostini (2005),  who discusses a development from the
standard formalism (see Kendall \& Stuart 1979) basing on Bayesian statistics.
This is an unbiased method to perform fits on data, including uncertainties on both axes
and a term of extravariance $\sigma_{v}$ given by the fluctuations of a``hidden"
parameter. The log-likelyhood function discussed by D'Agostini (2005) reads 
\begin{eqnarray}
L(m,q,\sigma_{v};\mathbf{x} ,\mathbf{y})&=&
\frac{1}{2} \sum_i
 \rm log(\sigma_{v}^2+\sigma^2_{y_i} +m^2 \sigma^2_{x_i})+
\nonumber\\
&&\frac{1}{2}\sum_i
 \frac{(y_i-m x_i-q)^2}{\sigma_{v}^2+\sigma^2_{y_i}
 +m^2 \sigma^2_{x_i}}~,
\end{eqnarray} on having  used two  models, namely:
\begin{equation}
y_{lin}=x_i\, m+q ~;
\end{equation}
and a log-linear model $Y = \log y,\, X = \log x$ with uncertainties
evaluated on using standard propagation to yield
\begin{equation}
Y=\log x_i\, m+q ~. 
\end{equation}
 This method was successfully used by Guidorzi et al. (2005). 
\begin{acknowledgements}
     We  thank S.  Bianchi, M.  Tomei,  M. Perri  for their help in the
     use of the \xmm~ Science  Analysis System  (SAS),  and  P. Giommi, E. Massaro, G.
     Tosti and A. Paggi, for useful discussions. One of us  (F. Massaro) acknowledges 
     support  by a fellowship of the Italian Space Agency (ASI), 
     in the context of the AGILE Space Mission.  We thank our referee for helpful
     comments  towards improving our presentation.
\end{acknowledgements}

\end{document}